\documentclass[fleqn,usenatbib]{mnras}

\usepackage{newtxtext,newtxmath}
\usepackage[T1]{fontenc}

\DeclareRobustCommand{\VAN}[3]{#2}
\let\VANthebibliography\thebibliography
\def\thebibliography{\DeclareRobustCommand{\VAN}[3]{##3}\VANthebibliography}

\usepackage{graphicx}	% Including figure files
\usepackage{amsmath}	% Advanced maths commands
\usepackage{lscape}
\usepackage{geometry}
\usepackage{comment}
\usepackage{gensymb}
\usepackage{float}
\newcommand{\Cl}{CL~J1449+0856}

\newcommand{\kms}{km s$^{-1}$}

\title[Multiple faint Radio-jets in \Cl]{Feedback Factory: Multiple faint radio-jets detected in a cluster at z=2}

\author[B. S. Kalita]{Boris S. Kalita,$^{1}$\thanks{E-mail: boris.kalita@cea.fr}
Emanuele Daddi,$^{1}$
Rosemary T. Coogan,$^{2}$
Ivan Delvecchio,$^{1,3}$
Raphael Gobat,$^{4}$
\newauthor 
Francesco Valentino,$^{5,6}$
Veronica Strazzullo,$^{7,8,3,9}$
Evangelia Tremou,$^{10}$
Carlos G\'omez-Guijarro,$^{1}$
\newauthor
David Elbaz,$^{1}$
and Alexis Finoguenov$^{11}$
\\ \\
$^{1}$CEA, Irfu, DAp, AIM, Universit\`e Paris-Saclay, Universit\`e de Paris, CNRS, F-91191 Gif-sur-Yvette, France \\
$^{2}$Max-Planck-Institut für Extraterrestrische Physik (MPE), Giessenbachstr.1, 85748 Garching, Germany\\
$^{3}$INAF—Osservatorio Astronomico di Brera, via Brera 28, I-20121, Milano, Italy\\
$^{4}$Instituto de Física, Pontificia Universidad Católica de Valparaíso, Casilla 4059, Valparaíso, Chile\\
$^{5}$Cosmic Dawn Center (DAWN), Copenhagen, Denmark\\
$^{6}$Niels Bohr Institute, University of Copenhagen, Jagtvej 128, DK-2200, Copenhagen, Denmark\\
$^{7}$Faculty of Physics, Ludwig-Maximilians-Universität, Scheinerstr. 1, 81679 Munich, Germany\\
$^{8}$University of Trieste, Piazzale Europa, 1, 34127 Trieste TS, Italy\\
$^{9}$INAF - Osservatorio Astronomico di Trieste, via Tiepolo 11, I-34131, Trieste, Italy\\
$^{10}$LESIA, Observatoire de Paris, CNRS, PSL, SU/UPD, Meudon, France\\
$^{11}$Department of Physics, University of Helsinki, Gustaf H{\"a}llstr{\"o}min katu 2a, FI-00014 Helsinki, Finland
}

\date{Accepted 2021 February 22. Received 2021 February 18; in original form 2020 December 4}

\pubyear{2020}

\begin{document}
\label{firstpage}
\pagerange{\pageref{firstpage}--\pageref{lastpage}}
\maketitle

\begin{abstract}
%This is a simple template for authors to write new MNRAS papers.
%The abstract should briefly describe the aims, methods, and main results of the paper.
%It should be a single paragraph not more than 250 words (200 words for Letters).
%No references should appear in the abstract.
We report the detection of multiple faint radio sources, that we identify as AGN-jets, within CLJ1449+0856 at z=2 using 3\,GHz VLA observations. We study the effects of radio-jet based kinetic feedback at high redshifts, which has been found to be crucial in low redshift clusters to explain the observed thermodynamic properties of their ICM. We investigate this interaction at an epoch featuring high levels of AGN activity and a transitional phase of ICM in regards to the likelihood of residual cold-gas accretion. We measure a total flux of $\rm 30.6 \pm 3.3~\mu Jy$ from the 6 detected jets. Their power contribution is estimated to be $1.2 ~(\pm 0.6)~ \times 10^{44} ~\rm ergs~ s^{-1}$, although this value could be up to $4.7 ~ \times 10^{44} ~\rm ergs~ s^{-1}$. This is a factor $\sim 0.25 $ -- $1.0$ of the previously estimated instantaneous energy injection into the ICM of CLJ1449+0856 from AGN outflows and star formation, that have already been found to be sufficient in globally offsetting the cooling flows in the cluster core. In line with the already detected abundance of star formation, this mode of feedback being distributed over multiple sites, contrary to a single central source observed at low redshifts, points to accretion of gas into the cluster centre.  This also suggests a `steady state' of the cluster featuring non cool-core like behaviour. Finally, we also examine the TIR-radio luminosity ratio for the known sample of galaxies within the cluster core and find that dense environments do not have any serious consequence on the compliance of galaxies to the IR-radio correlation.

\end{abstract}


%%%%%%%%%%%%%%%%%%%%%%%%%%%%%%%%%%%%%%%%%%%%%%%%%%

%%%%%%%%%%%%%%%%% BODY OF PAPER %%%%%%%%%%%%%%%%%%

\section{Introduction} \label{sec:intro}

Discoveries brought by steadily increasing, mainly X-ray observing capabilities have generated a debate over the heating and cooling processes shaping the intra-cluster medium (ICM) and, by extension, their respective clusters. If left unhindered, the gravitational collapse and the subsequent cooling via X-ray emission is known to produce a steady inflow of gas into cluster cores \citep[known as the classical cooling flow model;][]{fabian94}. This `cooling' is characterised by significant loss of energy in a very short time ($\ll \rm 1/H_{0}$). Although observed in moderate amounts, the high levels of cool gas in cluster cores expected from such a scenario have not been detected \citep[e.g.,][]{peterson01,peterson03, sanders08} nor has been the resulting star formation and CO emission \citep[e.g.,][]{mcnamara89,edge03}. Moreover, this would have also led to a galaxy population at the centre of clusters much more massive and brighter than what the well-established truncation of the high-luminosity end of the galaxy luminosity function allows \citep{benson03}. 

With overwhelming evidence for heating mechanisms to be in place to control the cooling flows and suppress the growth of giant elliptical galaxies in cluster cores, multiple modes of energy injection have been proposed over the years. While thermal conduction \citep[e.g.,][]{kim03a, pope06}, supernova explosions \citep{springel03} and turbulent mixing \citep[e.g.,][]{kim03b, voigt04, dennis05} can all be possible heating mechanisms, they have been found to be insufficient on their own \citep{kravtsov00,borgani04,voigt04}. However, AGN driven ICM heating provides an ideal self-sustained mechanism sufficient in providing enough entropy to prevent a runaway global cooling of the ICM \citep{mcnamara07} and in the process, regulate star formation \citep{voit15,voit17, tremblay18, olivares19, russell19}. In the nearby universe, the interaction of radio-loud AGN-jets with the X-ray emitting ICM of clusters has been well established using high quality data from Chandra and XMM-Newton \citep[][for a review]{mcnamara07, gitti12}. They revealed X-ray deficit cavities that were created due to the kinetic feedback from AGN-jets from the central brightest cluster galaxy (BCG) in the heart of cluster cores. \citep[e.g.,][]{boehringer93,mcnamara00,blanton01,mcnamara01,heinz02,clarke04}. These cavities generate shocks which dissipate the internal enthalpy into the surrounding medium in their wakes \citep{jones02, fabian03, birzan04}. The most typical configuration observed in low redshift clusters are of bipolar jet-like flow of radio-bright plasma manifesting in the form of expanding lobes, emanating from the central dominant (cD) galaxy at the cluster centre.

There has been so far no clear evidence for jet-driven feedback at $\rm z \ge 2$. It is noteworthy that this is a crucial epoch featuring a peak in AGN activity \citep[][for a review]{heckman14} along with the collapse of non-virialised `protoclusters' to form lower-redshift virialised clusters \citep{overzier16}. Although there have been studies on the feedback from powerful radio galaxies at high redshifts \citep[e.g.,][]{miley06, venemans07,nesvadba08, hatch09, nesvadba17, markov20} these are usually biased towards the high luminosity end and hence unlikely to be representative of a more general scenario. Complicating matters further, there have been suggestions that the energy required to maintain the ICM of a cluster in its observed thermodynamic form at low-redshifts may be different from that required to bring it to this configuration at the first place \citep{mccarthy08}. 

Keeping in mind the need to further investigate the $\rm z \ge 2$ epoch, we present our results from the study of \Cl\ at a redshift of 1.99 \citep{gobat11, gobat13}. We have undertaken a sensitive radio analysis of this high redshift X-ray detected cluster primarily aimed at uncovering feedback mechanisms in place which are likely to play a major role in the evolution of this cluster. Following up on the measurements in \citet{valentino16} for the AGN outflows from the two known X-ray cluster AGNs, we aim to address the AGN-jet contribution in this work.

Our paper is structured as follows: after introducing \Cl\ in Sec.~\ref{sec:clj1449}, we describe the observations and data analysis in Sec.~\ref{sec:obs}. Results regarding radio flux measurements are presented in Sec.~\ref{sec:results} while their consequences are discussed in Sec.~\ref{sec:discussion}. Finally, Sec.~\ref{sec:conclusion} with the conclusions brings this paper to a close. Throughout, we adopt the concordance $\Lambda$CDM cosmology, characterized by  $\Omega_{m}=0.3$, $\Omega_{\Lambda}=0.7$, and $H_{0}=70$ km s$^{-1}\rm Mpc^{-1}$. We use a Chabrier initial mass function \citep[IMF;][]{chabrier03}. All images are oriented such that north is up and east is to the left.

\section{CL J1449+0856} \label{sec:clj1449}

\begin{table*}
\centering
\begin{tabular}{cccccc@{\hspace*{1mm}}c@{\hspace*{1mm}}cc}
ID & RA & Dec & $z_{CO}$  & F$\rm _{3\,GHz}$ &  F$\rm _{870\mu m}$ & log(M$_\ast$/M$_{\odot}$) & SFR & Description \\ 
& (deg) & (deg) & & ($\rm \mu$Jy) & ($\rm \mu$Jy) & & (M$_{\odot}$ yr$^{-1}$) &\\ \hline
A2 & 222.30710 & 8.93951 &$\;1.9951\pm0.0004$ &$\;4.2 \pm 1.3$ & $\;515 \pm 135$ & $\;9.93^{**}$  & 94$\pm$16 & Merger \\
A1 & 222.30872 & 8.94037 &$\;1.9902\pm0.0005$ &$\;7.8 \pm 1.3$ & $\;1370 \pm 140$ & $\;10.28^{**}$  & 178$\pm$19 & Merger \\
13 & 222.30856 & 8.94199 &$\;1.9944\pm0.0006$ &$\;6.6 \pm 1.3$ & $\;248 \pm 69$ & $\;10.46 \pm 0.3$  & 39$\pm$8 & AGN \\ 
A6 & 222.30991 & 8.93779 &$\;1.9832\pm0.0007$ &$\;<2.7$ & $\;709 \pm 75$ & $\;10.71 \pm 0.3$  & 88$\pm$11 & Prominent bulge\\ 
N7 & 222.31029 & 8.93989 &$\;1.9965\pm0.0004$ &$\;<2.7$ & $\;217 \pm 79$ & $\;10.07 \pm 0.3$  & 31$\pm$10 & Interacting \\
B1 & 222.30891 & 8.94071 &$\;1.9883\pm0.0070$ &$\;4.9 \pm 1.3$ & $\;346 \pm 69$ & $\;10.81 \pm 0.3$  & 38$\pm$8 & Merger \\
3 & 222.30910 & 8.93951 &$\;1.9903\pm0.0004$ &$<2.7$ & $<141$ & $\;10.31 \pm 0.3$  & $23\pm9$ & Quiescent \\
S7 & 222.31021 & 8.93980 &$\;1.982\pm0.002^{*}$&$\;<2.7$ & $<150$ & $\;10.48 \pm 0.3$  & $\sim 8$ & AGN, interacting \\
2 & 222.30586 & 8.94297 &$\;1.98\pm0.02$ &$\;<2.7$ & $\;184 \pm 73$ & $\;10.81 \pm 0.3$  & $<24$ & Prominent bulge \\
H4 & 222.30859 & 8.94100 &$\;-$ &$\;<2.7$ & $\;<135$ & $\;10.9 \pm 0.3$  & $<22$ & Quiescent, merger \\ 
H5 & 222.30872 & 8.94064 &$\;-$ &$\;4.4 \pm 1.3$ & $\;237 \pm 95$ & $\;10.9 \pm 0.3$  & $<27$ & Quiescent, merger \\ \hline
\end{tabular}
\caption{The confirmed cluster members of \Cl. The first seven have molecular gas detection as reported in \citet{coogan18} while 2, H4 and H5, which only have photometric redshift confirmations,  are additions from \citet{gobat11, valentino15, strazzullo16, strazzullo18}. $^{*}$ S7 lacks molecular gas detection, and hence the optical redshift calculated using \textit{HST/WFC3} grism and \textit{MOIRCS} spectroscopy \citep{gobat11, valentino15} has been reported. $2\sigma$ upper limits are reported for galaxies lacking F$\rm _{3\,GHz}$ and/or F$\rm _{870\mu m}$. The star formation rates (SFR) reported are the average of the estimates from CO[4--3] line flux and the $\rm {870\mu m}$ continuum flux reported in \citet{coogan18}. \textit{($^{**}$ Stellar mass measurements from dynamical mass estimates \citep{coogan18})}
\label{tab:galz}}
\end{table*}

\begin{figure*}
    \centering
    \includegraphics[width=0.95\textwidth]{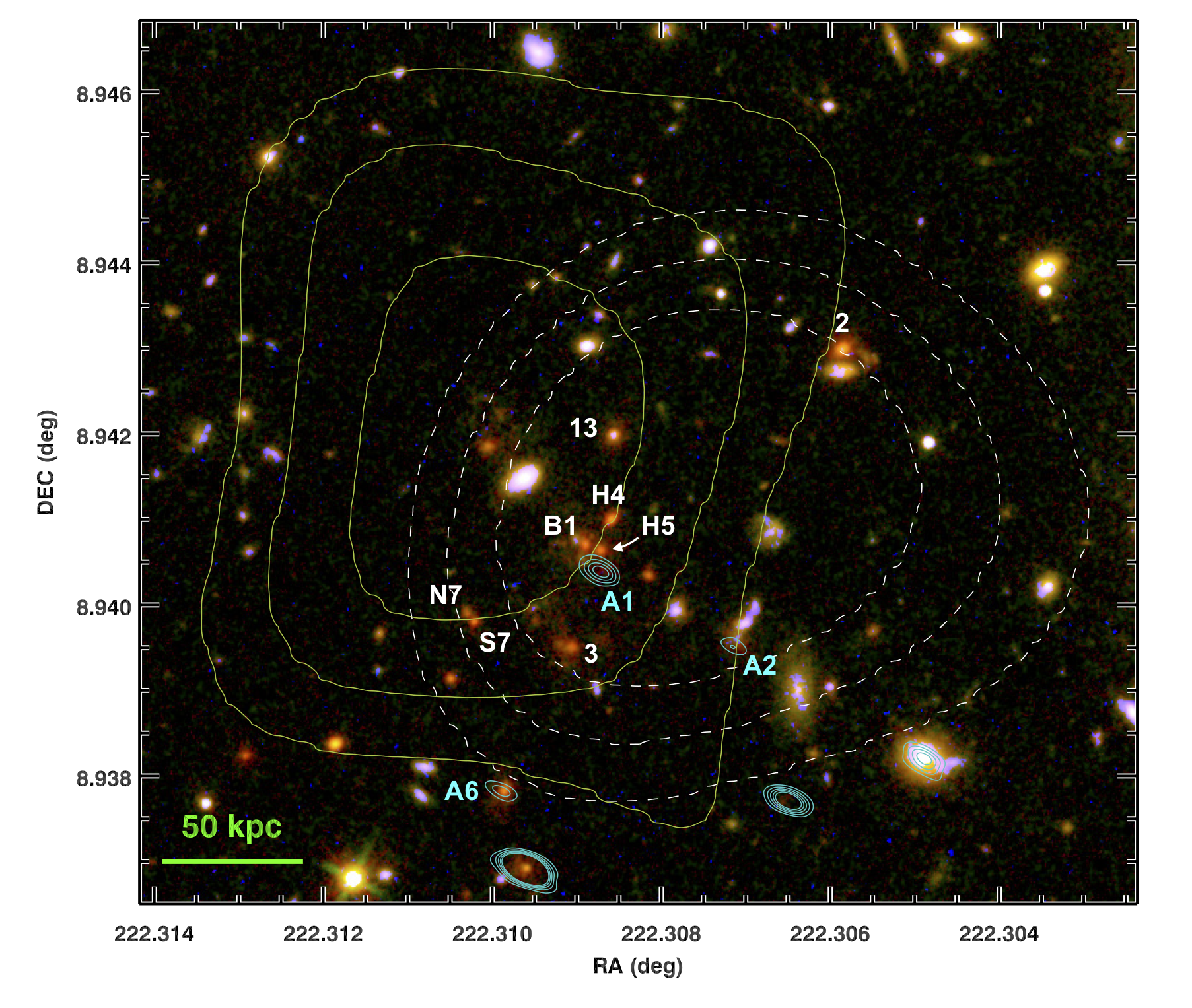}
    \caption{\textit{HST/WFC3} color composite image of CL J1449+0856. The grey dashed contours represent the X-ray intensity map from Chandra \citep{valentino16} while the extended green contours display the SZ signal tracing the mass distribution \citep{gobat19}. The cyan contours are the 870 $\mu$m ALMA sources. All galaxies listed in Table~\ref{tab:galz} have been marked.}
     \label{fig:clj_full}
\end{figure*}

CL J1449+0856 (Fig.~\ref{fig:clj_full}), is in a phase of galaxy assembly and star formation quenching which will eventually lead to the creation of the dominant population of massive and passive galaxies that characterizes later galaxy clusters \citep{strazzullo16}. Unlike the less evolved protoclusters \citep[][for a review]{overzier16} that are more commonly found at z $\ge$ 2, \Cl\ features an extended X-ray emission originating most likely from its hot ICM plasma  \citep[as confirmed now by an SZ detection;][]{gobat19} along with an already forming red sequence \citep{strazzullo16}. The observed X-ray emission is a signature of a dense ICM environment, similar to what is observed in local galaxy clusters. It also hosts a population of highly star-forming galaxies which suggests a presence of large scale-gas inflow needed for their sustenance \citep{valentino15}. We list the various galaxies with sub-mm continuum and/or CO line detection \citep{coogan18,strazzullo16} within the core of \Cl\ in Table~\ref{tab:galz}. These will be the galaxies that we will be restricting our work to (in the region r $<$ 200 kpc from the cluster center). However, two additional near-infrared (NIR) detected cluster members identified in the \textit{HST/WFC3} imaging but undetected in the sub-mm observations are also included in this table and the analysis. We now discuss the primary galaxy populations in \Cl\ which we shall be extensively referring to in the following sections:
\begin{enumerate}
    \item \textbf{The assembling BCG:} The two galaxies B1 and H5, detected in NIR and radio continuum, with a third (A1) detected in sub-mm and radio continuum, are expected to contribute to the formation of the future BCG of \Cl. \citet{coogan18} present a faint detection of the CO[4-3] line for B1 (3.7$\sigma$) and a 870$\mu$m continuum flux of 346$\pm$69 $\mu$Jy  whereas H5 has only been marginally detected ($\sim 2.5 \sigma$) at 870$\mu$m. Conversely for A1, the CO[4-3] line has been detected at 11.9$\sigma$ and its 870$\mu$m continuum flux has been found to be 1370$\pm$140 $\mu$Jy. Regarding the dynamics, both \citet{coogan18} and \citet{strazzullo18} conclude that B1, H5 and A1 \citep[along with H4;][]{strazzullo18} are undergoing a merger at the cluster center. Moreover, B1 and H5 (along with H4) appear to be likely suppressed/quiescent galaxies \citep[lower main-sequence galaxies;][]{sargent14} while A1 has a high SFR of 177.5$\pm$19 M$_{\odot}$ yr$^{-1}$, above what is expected from a main sequence galaxy \citep[Fig. 8 in ][]{coogan18}.      
    \item \textbf{Merging galaxies with X-ray AGN:} One of the two X-ray detected AGNs, S7, originally reported by \citet{brusa05} and confirmed by \citet{campisi09}, is also part of a merging pair with galaxy N7 \citep[with a velocity separation of $\sim$1380 \kms at a projected separation $<$0.5$^{\prime\prime}$ or $\sim$4 kpc;][]{coogan18}. Although the N7-S7 system has a combined 870$\mu$m flux measurement of 217$\pm$79 $\mu$Jy, it can mainly be associated with N7. This leaves S7 with a very low SFR ($\sim$8 M$_{\odot}$ yr$^{-1}$, based on a negligible CO[4-3] detection), which could be a result of its AGN activity \citep[with an L$_{x}=10^{43.7}$ ergs s$^{-1}$;][]{campisi09} suppressing star formation \citep{barger15}. 
    \item \textbf{Isolated X-ray AGN} The second X-ray detected cluster AGN: 13, like S7, was also reported and confirmed by \citet{brusa05} and \citet{campisi09}. However unlike S7, it is isolated and features an appreciable 870$\mu$m flux of 248$\pm$69 $\mu$Jy, placing it comfortably among the lower main-sequence star formation suppressed galaxies \citep{sargent14}. 

\end{enumerate}

Furthermore, there is a vast Lyman-$\alpha$ halo \citep[$\ge$ 100 kpc,][]{valentino16} in the core of \Cl, possibly being powered by outflows from the two cluster AGNs along with the SFR in the cluster core \citep[with an estimated net energy outflow rate, $\dot E_{kin} \sim 5 \times 10^{44}$ ergs s$^{-1}$;][]{valentino16}. This is 5 times larger than the Lyman-$\alpha$ extended luminosity, thereby opening up a possibility that the rest is being injected into the ICM in order to offset the global radiative cooling in the cluster ICM.

\section{Data observation and reduction} \label{sec:obs}
Our analysis is based on a multi-wavelength dataset that we present below. We combine 3\,GHz Very Large Array (VLA) continuum data with \textit{ALMA} sub-mm and \textit{HST/WFC3} NIR observations to extract a complete picture of galaxies irrespective of their levels of dust obscuration. The corresponding flux measurements along with resulting star formation rates (SFR) and stellar mass (M$_{\ast}$) acquired from this data have already been presented in ~\citep{coogan18} and have been reproduced in table~\ref{tab:galz}. However, we revise the 3\,GHz flux measurements in column 5 following the re-analysis of the corresponding data, which have been the primary motivation for this work. Furthermore, additional previously unreported low resolution GMRT 325MHz legacy continuum observations are also presented in this work, mainly providing upper limits.     

\subsection{JVLA S-band} \label{sec:vla_obs}
Continuum observations centered at 3 GHz were obtained for the cluster using the Karl G. Jansky Very Large Array (JVLA; project code: 12A-188, PI: V. Strazzullo). These observations were carried out between February and November 2012, for a total on-source time of $\sim 1.1$h in configuration C and $\sim 11.4$h in configuration A. Quasar J1331+3030 was used for flux calibration in both cases. All data were calibrated and imaged following the standard procedures with the Common Astronomy Software Application \citep[CASA; ][]{mcmullin07}. Additional flagging had to be done especially in spectral windows 13, 14 and 15 for low level radio-frequency interference (RFI) which significantly affected the phase of most of the data in both configurations. For cleaning the A-configuration calibrated data, we used the TCLEAN task with a Briggs weighting scheme featuring a robust parameter of 2.0. This was aimed at facilitating the highest sensitivity possible for source detection. However, we also created images with lower sensitivities and therefore higher resolutions, in order to examine any undetected compact sources. We found none. For the masking of sources, we used the `auto-threshold' feature in tandem with manual selections. Frequency dependent clean components (with two Taylor terms; \verb|nterms=2|) were also used in imaging to mitigate large-bandwidth effects \citep{rau11}. Additionally, we implemented widefield imaging with 128 projection planes (\texttt{wprojplanes=128}) and image deconvolution was carried out with the multi-scale multi-frequency synthesis algorithm on scales of [2,5,10] pixels \citep{rau11} for improved handling of the extended sources present in the field. A similar procedure was used for C-configuration, however with a robust parameter of 0.5 in order to achieve a balance between the resolution and sensitivity. Moreover, we used the mask created during the A-configuration analysis as a starting point for the masking of strong sources. This was then modified incrementally through a mix of automated and manual cleaning. The primary beam FWHM of the 3 GHz observations was 15', and the FWHM of the synthesised beam at this wavelength was $\sim0.83^{\prime\prime}\times0.67^{\prime\prime}$ at PA = -4.17° for the A-configuration data, with an RMS noise of 1.34$\mu$Jy/beam over an effective bandwidth of 1.5 GHz. In the C-configuration, the synthesised beam was $\sim7.44^{\prime\prime}\times6.58^{\prime\prime}$ at PA = 7.80°, with an RMS noise of 12.1 $\mu$Jy/beam over a similar effective bandwidth.

\subsection{GMRT P-band} \label{sec:gmrt_obs}

Low-frequency radio continuum observations were done using the Giant Metrewave Radio Telescope (GMRT; project code: 23\_068, PI: R. Gobat) centered at 325 MHz on 23 May 2013. The bandwidth of these observations was 32 MHz and the total on-source time $\sim 4$ hours. These data were also reduced using CASA. We executed manual calibration and RFI flagging before initiating the cleaning procedure with the TCLEAN task. In order to achieve the best combination of sensitivity and resolution possible from the data, we used the Briggs weighting scheme with a robust parameter of 0.5. The `auto-masking' feature was implemented to handle the abundance of bright sources within the field of view. We also executed multiple rounds of phase and amplitude self-calibration due to the presence of a S/N (signal-to-noise) $\sim 10^3$ source within the primary beam. The final synthesised beam was $\sim 7.4^{\prime\prime} \times 6.6^{\prime\prime}$ at PA = 7.8° and the RMS noise of the resulting image was 92 $\mu$Jy/beam.

\subsection{Sub-millimeter} \label{sec:fir_obs}

Atacama Large Millimetre Array (ALMA) band 4 observations of the cluster were taken in Cycle 3 (Project ID: 2015.1.01355.S, PI: V. Strazzullo) while band 3 and 7 were taken in Cycle 1 (Project ID: 2012.1.00885.S, PI: V.~Strazzullo). The details of these observations and the corresponding analysis have already been discussed in \citet{coogan18}. We borrow their results from the spectral line flux extraction of CO[4-3] and CO[3-2] along with the continuum flux measurements at 870\,$\mu$m obtained for each of the cluster galaxies listed in Table~\ref{tab:galz}. We also use the corresponding 870\,$\mu$m continuum image with a sensitivity of 67.6\,$\mu$Jy/beam. We direct the reader to \citet{coogan18} for the details of the data reduction and analysis. 

Furthermore, ALMA+ACA data of \Cl\ at 92 GHz was presented in \citet{gobat19} in cycle 4 (Project ID: 2016.1.01107, PI: R.~Gobat). We here make use of their results, including the Sunyaev-Zel’dovich effect (SZ) 5$\sigma$ signal and the corresponding residual map.

\subsection{X-ray: soft-band}
\citet{brusa05} and \citet{campisi09} presented deep X-ray soft-band (0.5--2 keV) with \textit{XMM-Newton} and \textit{Chandra} telescopes respectively totalling 80 ks each. After subtraction of the point sources associated with the cluster AGNs detected using \textit{Chandra}, \citet{gobat11} detected diffuse X-ray emission in the \textit{XMM-Newton} data for \Cl\ at $3.5\sigma$ on scales of 20--30$^{\prime\prime}$. \citet{valentino16} however got an improvement on this having detected the same signal at $4\sigma$ with additional \textit{Chandra} data with nominal exposure of 94.81 ks. We shall be henceforth using this result in our work.

\section{Results} \label{sec:results}

\begin{figure}
    \centering
    \hspace*{-1mm}\includegraphics[width=0.48\textwidth]{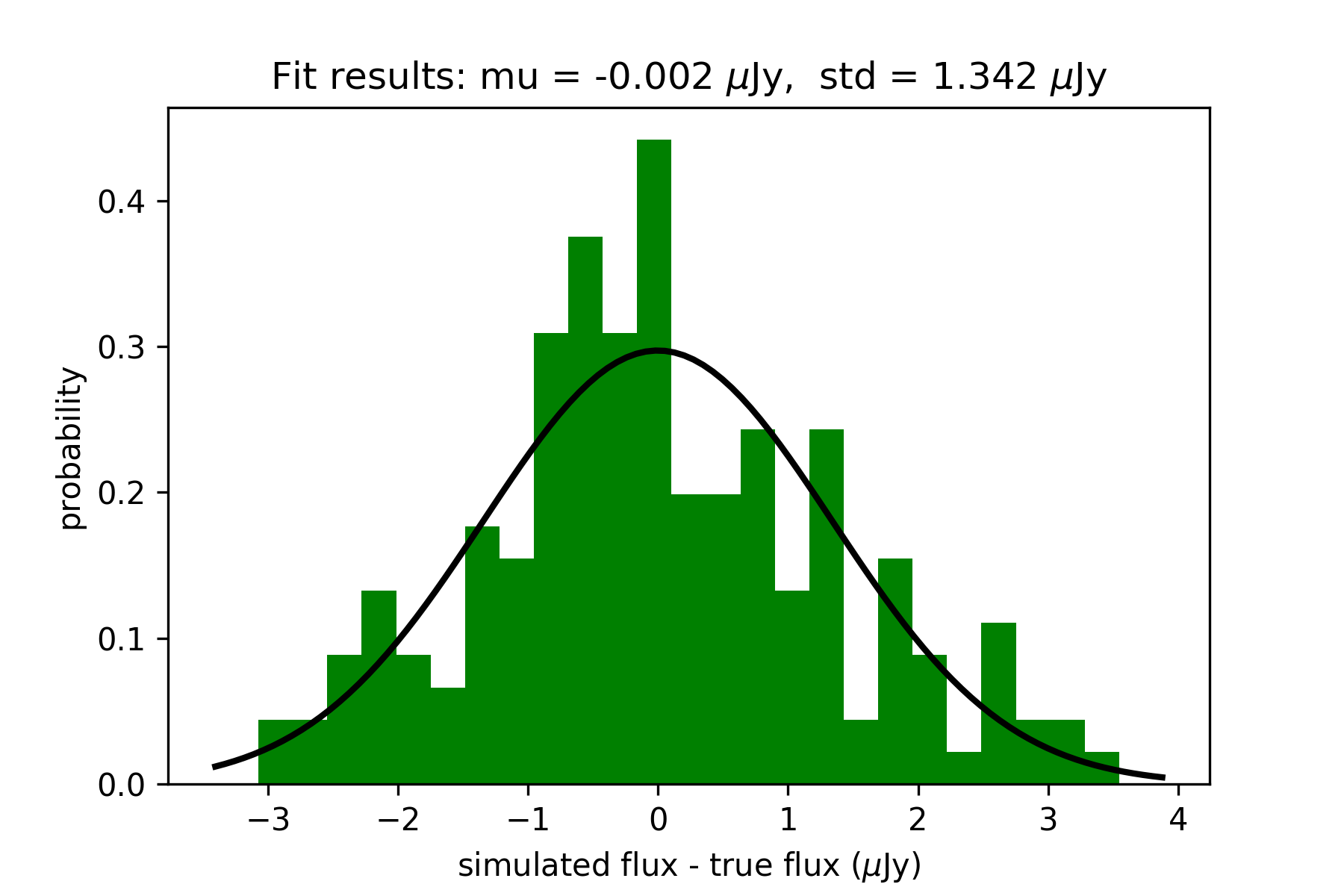}
    \caption{The histogram created out of the source detection simulation carried out within a distance of $\sim 80^{\prime\prime}$ from the centre of \Cl\ on the VLA 3\,GHz image. The x-axis gives the difference between the measured flux of the simulated sources and their true values, while the y-axis provides a measure of the fractional occurrence. We fit a gaussian, the $\sigma$ of which is the RMS noise/beam of in the data}
    \label{fig:err_est}
\end{figure}

\begin{figure*}
    \centering
    \includegraphics[width=0.95\textwidth]{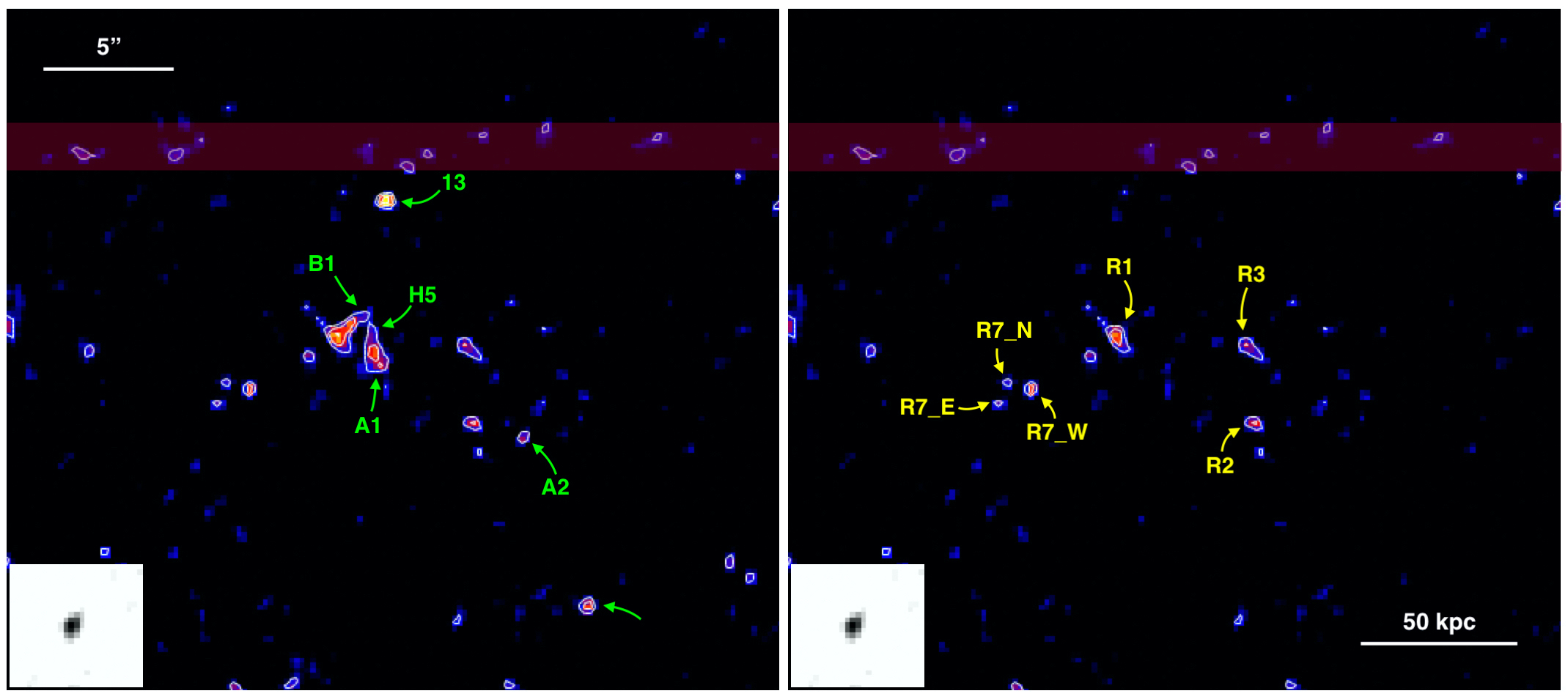}
    \caption{\textbf{(Left)} The VLA 3\,GHz image of the cluster core with the galaxies in Table~\ref{tab:galz} detected ($> 3 \sigma$) with an additional bright interloper and \textbf{(right)} the subsequent residual image after their subtraction, with the remaining major emission regions marked. The sigma contours in white are also provided for each of the images, starting from 3$\sigma$ with increments of 1$\sigma$. Finally, the white boxes show the VLA 3\,GHz PSF in the same spatial scale as that in the images. The maroon shaded region highlights a very faint imaging residual spike from the strongest source in the VLA field of view, which contributes to a few spurious $\sim 3 \sigma$ peaks. No such systematics are present in the rest of the image.}
     \label{fig:vla_1_2}
\end{figure*}

\subsection{Radio continuum flux measurement} \label{sec:radiores}
 
 \begin{table*}
\centering
\begin{tabular}{cccc@{\hspace*{1mm}}c@{\hspace*{1mm}}c@{\hspace*{1mm}}c}
ID & RA & Dec & F$\rm _{3\,GHz}$ &  F$\rm _{870\mu m}$ & q$\rm _{TIR}$ & Possible galaxy \\ 
& (deg) & (deg) & ($\rm \mu$Jy) & (expected; $\rm \mu$Jy) & & association\\ \hline
R1 & 222.30911 & 8.94060 &$\;7.4\pm1.3$ &$\;570\pm130$ & $<1.8$ & B1, H5, A1 \\
R2 & 222.30765 & 8.93966 &$\;6.2\pm1.3$ &$\;480\pm130$ & $<1.9$ &uncertain \\ 
R3 & 222.30775 & 8.94050 &$\;5.6\pm1.3$ &$\;440\pm130$ & $<2.0$ &uncertain \\
R7\_N & 222.31031 & 8.94012 &$\;3.4\pm1.3$ &$\;260\pm130$ & $<2.1$ &S7, N7 \\
R7\_E & 222.31041 & 8.93988 &$\;3.4\pm1.3$ &$\;270\pm130$ & $<2.1$ &S7, N7 \\
R7\_W & 222.31006 & 8.94004 &$\;4.6\pm1.3$ &$\;360\pm130$ & $<2.0$ &S7, N7 \\ \hline
\end{tabular}
\caption{The radio jet detections with their observed 3\,GHz flux along with that expected at 870\,$\mu$m due to star formation if these sources would have been MS galaxies. The penultimate column gives the $3\sigma$ upper limits of the $\rm q_{TIR}$. The final column lists the possible source galaxies from Table~\ref{tab:galz} for each of the jet sources, however such associations have not been possible for R2 and R3.   
\label{tab:jets}}
\end{table*}
 
We first estimate the sensitivity reached in the VLA 3\,GHz A-configuration, which we mainly use for this analysis. We use the software GALFIT \citep{peng10} to artificially place point sources (about 200 in number) in an area of $\sim 80^{\prime\prime}$ in radius centered at the location of the cluster centre. The flux of these sources is randomly chosen within the range 4.5\,$\mu$Jy -- 7.5\,$\mu$Jy which is where most of the radio-detected cluster galaxies lie \citep{coogan18}. Regions that are within a distance of 1$^{\prime\prime}$ from any $> 4\sigma$ pixel have been ignored during this exercise in order to prevent source confusion. GALFIT is used again on the resulting image to fit these sources with fixed positional priors. The resulting $1\sigma$ error in the point-source sensitivity is hence found to be 1.34 $\mu$Jy (Fig.~\ref{fig:err_est}), which is a 34\% improvement compared to that in \citet{coogan18}. This can be attributed to the additional flagging that we implemented that would lead to an improvement in the phase solutions and thereby the imaging. Moreover, we used cleaning input parameters more in line with our primary goal of source detection, including a broader synthesised beam. The value 1.34\,$\mu$Jy is also equal to the RMS noise/beam of the 3\,GHz image at phase centre (Sec.~\ref{sec:vla_obs}). We shall therefore be using the terms `point-source sensitivity' and `RMS noise/beam' interchangeably for the rest of the paper.

We then measure the 3\,GHz continuum fluxes by placing point sources with positional priors from an available \textit{HST/WFC3} image (and in case of optically dark galaxies, the 870 $\mu$m continuum ALMA data) for the galaxies listed in Table~\ref{tab:galz} and shown in Fig.~\ref{fig:clj_full}. We choose to use point sources since the size of the galaxies are smaller than the synthesised beam half-power beam width. However, we find that there are multiple additional sources ($\ge 3.5\sigma$) that remain after the fitting and subtraction of the known galaxies at the cluster center (Fig.~\ref{fig:vla_1_2}). Hence we execute another round of fitting with additional point sources having both position and flux free, besides the ones for the galaxies. The list of sources being fit is incrementally expanded until there are no remaining detectable sources present within the core of \Cl. We use the results from the final round of fitting where we incorporate the complete list of sources simultaneously. The resulting flux measurements of the galaxies are listed in Table~\ref{tab:galz}, while the additional sources, discussed in detail in the next section, have been recorded in Table~\ref{tab:jets}. It is noteworthy that we also fit point sources for two additional sources (R7\_N and R7\_E) that are at $\le 3.5\sigma$. This is due to their physical proximity to R7\_W and the AGN S7, details of which are discussed in Sec.~\ref{sec:disc_origin}. Following the final subtraction, we check for any residual emission around the fits. Finding none, we also conclude that all the sources detected are indeed unresolved point sources for the 3\,GHz synthesized beam.

We also detect an unresolved emission centered at the region with the assembling BCG and R1 with $\sim 3.8 \sigma$ significance in the C-configuration VLA observation. Using the A-configuration detections as positional priors, we attempt to measure individual flux of the sources. However, limited by the high levels of mixing of source fluxes due to a much lower resolution and a noise RMS higher by a factor of $\sim 10$ compared to the A-configuration image, we simply measure an integrated flux over the aforementioned region. We then use this flux as a constraint for the A-configuration flux measurements. We encounter a similar situation with our new 325\,MHz GMRT observations (Sec.~\ref{sec:gmrt_obs}), with a synthesized beam size $\sim7^{\prime\prime}$ and an RMS of 92 $\mu$Jy/beam, although it also lacks a $>3\sigma$ detection. We hence use this data to measure an upper limit value of $\sim 1.5$ for the negative radio spectral index of the sources detected using VLA. We apply this in the estimation of the uncertainty in luminosity measurements in the following sections.

\subsection{TIR-radio correlation} \label{sec:qtir_cal}

\begin{figure}
    \centering
    \hspace*{-1mm}\includegraphics[width=0.48\textwidth]{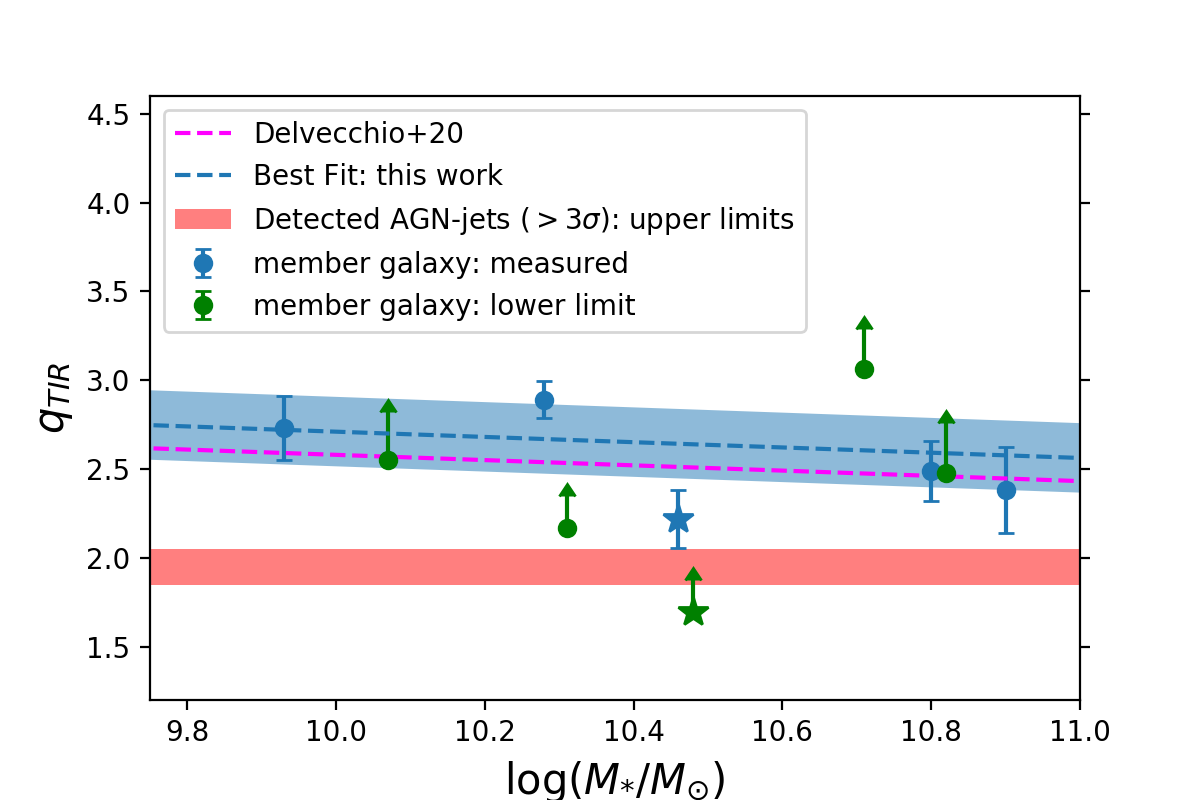}
    \caption{The total infrared and radio luminosity ratio ($\rm q_{TIR}$) vs the stellar mass of the member galaxies of \Cl\ (listed in Table.~\ref{tab:galz}), with the two AGNs shown with the `$\star$' marker. We also plot the best fit along with the 1$\sigma$ limits. We only provide the lower limit $\rm q_{TIR}$ for the galaxies lacking a radio detection which are obtained using the $2\sigma$ upper limits of the 3\,GHz VLA data. The slope was fixed to the value determined in \citet{delvecchio20}, the $\rm q_{TIR}$ vs. M$_{*}$ relation from which has also been presented for comparison. Finally, the $\rm q_{TIR}$ upper limits of the AGN radio-jets detected ($>3\sigma$) in this work has also been shown.}
    \label{fig:qFIR_M}
\end{figure}

Using the revised 3\,GHz flux values of the cluster core galaxies along with the total-infrared luminosity (L$_{\rm IR}$; rest-frame 8--1000 $\mu$m) that was derived from the 870\,$\mu$m flux presented in \citet{coogan18} using SEDs from \citet{bethermin15}, we measure the total-infrared (TIR) and radio luminosity ratio (q$_{\rm TIR}$) using the following relation \citep{helou88}:
\begin{center}
\begin{equation}
q_{\rm TIR}={\rm log}\left(\frac{L_{{\rm IR}}[{\rm W}]}{3.75\times10^{12}\rm Hz}\right)-{\rm log}\left(L_{1.4\,{\rm GHz}}[{\rm W\,Hz^{-1}}]\right),
\end{equation}
\end{center}
where $L_{1.4\,{\rm GHz}}$(; W\,Hz$^{-1}$) is the 1.4 GHz rest-frame luminosity measured from the 3\,GHz fluxes ($S_{3{\rm GHz}}$; W\,Hz$^{-1}$\,m$^{-2}$) using: 

\begin{center}
\begin{equation}
L_{1.4{\rm GHz}}=\frac{4\pi D_{L}^{2}}{(1+z)^{\alpha+1}}\left(\frac{1.4}{3}\right)^{\alpha}S_{3{\rm GHz}},\label{eq:Lrad}
\end{equation}
\par\end{center}
where D$_{L}$ is the luminosity distance and $\alpha$ is the spectral index. Since in this section we are targeting emission due to star formation, we assume an $\alpha=-0.7$ \citep{delhaize17}. We do note however, this value of $\alpha$ is not applicable for all kinds of objects. Additionally, \citet{coogan18} presented the stellar mass (M$_*$) of all but two galaxies (H4 and H5) listed in Table~\ref{tab:galz} that were measured from their respective SEDs. For A1 and A2, this was rather done using their dynamical masses \citep[][, for further details]{coogan18}. We use these values to plot q$_{\rm TIR}$ vs. M$_*$ (Fig.~\ref{fig:qFIR_M}). For those galaxies that are not detected at 870\,$\mu$m, we use the 2$\sigma$ limit as the flux in the measurement of L$_{\rm IR}$.

We also add to this the results from \citet{delvecchio20}, which is a follow-up study of Main-Sequence star-forming galaxies from \citet{delvecchio17} but for an $\rm M_{*}$-selected sample from the VLA-COSMOS 3\,GHz survey \citep{smolcic17}. Their work presents a q$_{\rm TIR}$ vs $\rm M_{*}$ relation as follows:
\begin{center}
\begin{equation*}
\rm q_{TIR}\left(M_{*},z\right) = 2.646\pm0.024 \cdot A^{\left(-0.023\pm0.008\right)} - B \cdot \left(0.148\pm0.013\right) ,\label{eq:qtir_M}
\end{equation*}
\par\end{center}
where A = (1+z) and B = (log $\rm M_{*}/M_{\odot}$ - 10). Since we just have 10 galaxies to work with, we only fit for the normalisation of the equation for a comparison of our sample with their results. As shown in Fig.~\ref{fig:qFIR_M}, we see an agreement within $1\sigma$ (0.2 dex), although acknowledging the inability of our limited dataset to detect a more subtle digression in the behaviour of q$_{\rm TIR}$, if it exists.

Furthermore, we measure the q$_{\rm TIR}$ $3\sigma$ upper limits (due to a lack of sub-mm detection) for the additional radio sources reported in the previous section and show their range on the y-axis of Fig.~\ref{fig:qFIR_M}. It is immediately clear that these limits are well below q$_{\rm TIR}$ of the galaxies listed in Table.~\ref{tab:galz}, as well as the those expected from the relation derived by \citet{delvecchio20} for main-sequence star-forming galaxies. The average q$_{\rm TIR}$ upper limit for the brightest of the additional sources (ignoring R7\_N and R7\_E due to <\,3$\sigma$ radio detection significance) is 1.95, in contrast with the average q$_{\rm TIR}$ of the 3\,GHz-detected galaxies, which is $\sim$ 2.5 as expected \citep{magnelli15}. This places the radio sources at least 0.5 dex below the general behaviour of galaxies within \Cl, making it unlikely that their radio emission is due to star formation.

\section{Discussion} \label{sec:discussion}

\subsection{Origin of the radio emissions} \label{sec:disc_origin}

\begin{figure*}
    \centering
    \includegraphics[width=0.966\textwidth]{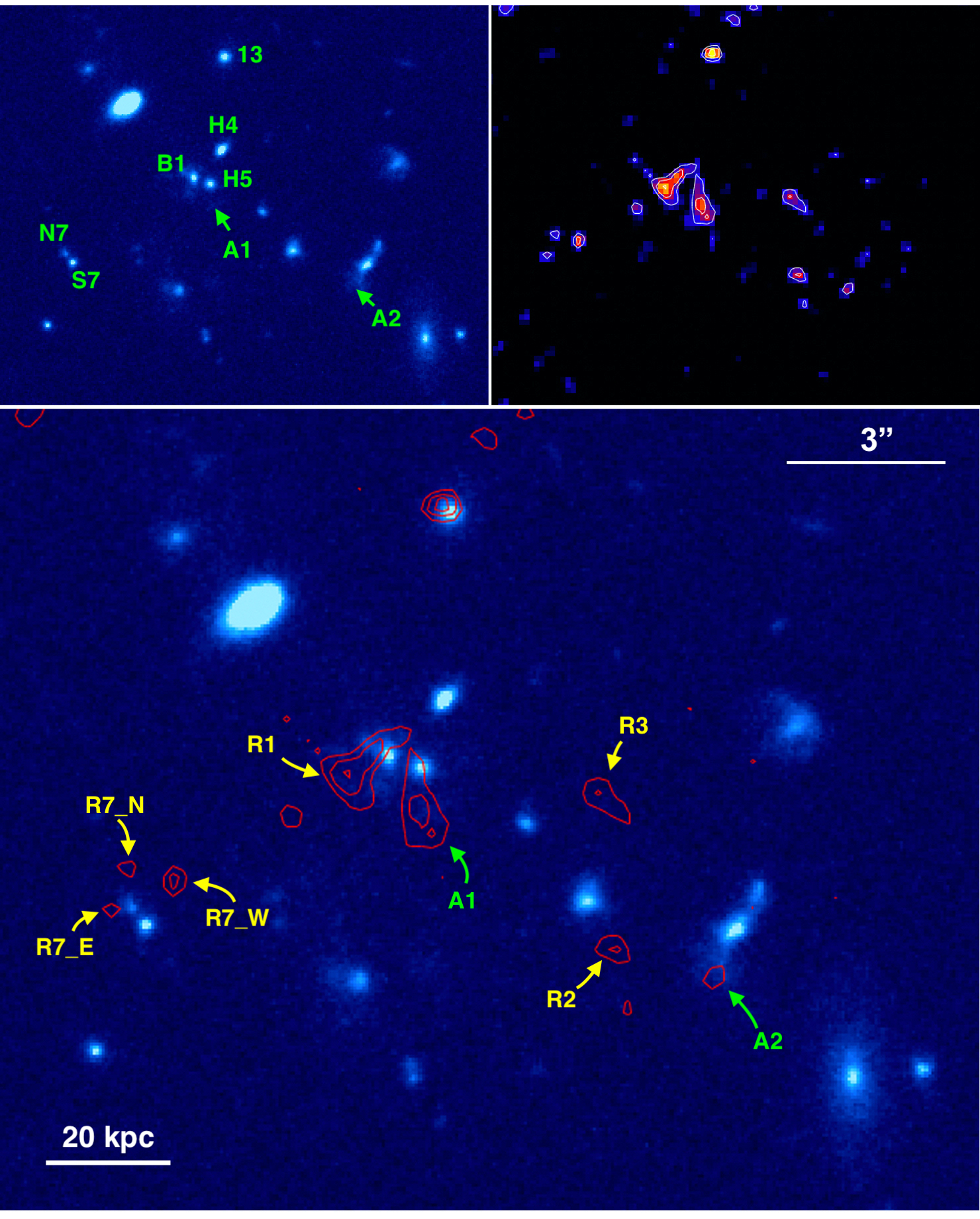}
    \caption{\textbf{(Top-left)} The \textit{HST/WFC3} F140W image with the sources listed in Table~\ref{tab:galz} marked. However, 2 and A6 are not shown as they are outside the area shown and are not relevant to the results presented in this work. \textbf{(Top-right)} The 3\,GHz VLA image of the region with sigma contours starting at 3$\sigma$ with 1$\sigma$ increments, as in Fig.~\ref{fig:vla_1_2}. \textbf{(Bottom)} The same \textit{HST/WFC3} F140W image with the contours of the VLA 3\,GHz image along with the AGN-jets listed in Table~\ref{tab:jets} (in yellow) and the two \textit{HST} undetected bright ALMA sources A1 and A2 that are also detected at 3\,GHz (in green) marked.}
     \label{fig:clj_core}
\end{figure*}

The detection of multiple emission regions that do not overlap with any of the HST or ALMA detected galaxies in the field (Figs.~\ref{fig:vla_1_2}, \ref{fig:clj_core}) raises the question of their authenticity and physical origin. To check the former, following \citet{jin19} we estimate the probability of these sources being due to noise fluctuations as: 
\begin{equation*}
    \rm P=1-\left(\rm P_{o}\right)^{n}
\end{equation*}
where $\rm P_{o}$ is the probability of finding a point source of a certain sigma within a Gaussian distribution while n is the number of synthesised beams that make up a region with radius 20$^{\prime\prime}$ (or $\sim$ 170 kpc) surrounding the centre of \Cl. This area was chosen for being large enough to cover the complete central core of the cluster. The probabilities for the three primary sources (R1, R2 and R3) are found to vary from $5 \times 10^{-5}$ for R1 to 0.03 for R3. It is noteworthy that these are generous upper limits, having not considered their relative physical proximity to each other and to the known galaxies of \Cl. This is however especially important in case of the R7 group (R7\_N, R7\_W and R7\_E) which appear clumped together and adjacent to a known cluster AGN, S7. Hence the possibility of these radio sources being generated due to noise is rather small.

We then move on to the investigation of their possible physical origins. To examine if these are simply undetected galaxies, which would mean that the radio flux could be traced back to star formation. We estimate an expected 870\,$\mu$m flux for each of them. This is done by first converting the measured 3\,GHz fluxes  ($S_{3{\rm GHz}}$; W\,Hz$^{-1}$\,m$^{-2}$)
into observed frame 1.4\,GHz ($S_{1.4{\rm GHz}}$; W\,Hz$^{-1}$\,m$^{-2}$), using a spectral index of --0.7 \citep[as expected from star formation;][]{delhaize17}. This is then converted to the expected flux at 870\,$\mu$m using SEDs from \citet{bethermin12}. From the resulting values listed in column 5 of Table~\ref{tab:jets} and the RMS noise of 67.5\,$\mu$Jy/beam in the 870\,$\mu$m data (Sec.~\ref{sec:fir_obs}), it is evident that these sources would have been comfortably detected at $>4\sigma$ had their emission been due to star formation. Moreover, the measurement of the 870\,$\mu$m expected flux for the sources is influenced by the assumption of a constant $\rm q_{TIR}$ \citep[2.64;][]{bethermin12}. Taking into account redshift evolution \citep{magnelli15, delhaize17} as well as stellar mass dependence \citep{delvecchio20} of $\rm q_{TIR}$, we expect even higher sub-mm fluxes, making our estimates much more conservative. It is noteworthy that this lack of detected flux has already manifested as the low q$_{\rm TIR}$ upper limits for the radio sources in Sec.~\ref{sec:qtir_cal}.

We also median stack 1.5$^{\prime\prime}$ cutouts of the 870\,$\mu$m continuum image from the regions of the sources and do not detect any emission even with the resulting sensitivity of 28\,$\mu$Jy/beam. This gives an 3$\sigma$ upper limit on the average SFR of $\sim 22~\rm M_{\odot} yr^{-1}$ within the emission regions, assuming a main-sequence template. With a similar stacking analysis on our HST F140W image we also confirm a lack of stellar presence down to a generous upper limit of AB magnitude of $10^{9}~\rm M_{\odot}$ \citep[for an AB magnitude of 25.9, assuming quiescence;][]{strazzullo16}. This eliminates a likelihood of these being galaxy merger driven tidal features, since had this been the case, there would have been structures with either detectable star formation or unobscured stellar emission. This only leaves two likely candidates: AGN radio-jets or cluster merger driven diffuse emission (radio halos and relics), both resulting from synchrotron emission. Given the concentrated nature of the emission sites, we can reject the latter. During the flux measurements (Sec.~\ref{sec:radiores}), we were able to account for all the detectable flux using PSF fitting, which suggests that the extent of the emission regions are within the A-configuration synthesised beam size of VLA at 3\,GHz ($\rm FWHM \sim 0.7^{\prime\prime}$ or 6 kpc). Moreover, most of these sites are spatially adjacent to either the assembling BCG members (B1, H5 and A1) or the south-eastern X-ray detected cluster AGN, S7 (Fig.~\ref{fig:clj_core}), hence pointing towards the radio-jet case.

We note our reliance on the positional accuracy of our VLA image with respect to the \textit{HST} astrometry. To check for any intrinsic offset between the two, we select all sources detected with $\ge 10 \sigma$ significance in VLA and also seen in the \textit{HST/WFC3} F140W image, over a range of distances from the radio-jet sites ($15^{\prime\prime}$--$70^{\prime\prime}$). We find nine such objects, which we fit with PSFs/Gaussians to get their positions in each of the images, separately. These are always found to be in agreement for each source within their respective intrinsic positional uncertainty in the VLA image, which can be estimated as $\rm \frac{1}{2} \frac{FWHM}{\left(S/N\right)}$ \citep{ball75,condon97}. We also repeat the same exercise with respect to the ALMA 870\,$\mu$m image and draw the same conclusion, although with only 4 sources that are detected at $\gtrsim 6 \sigma$ at 3\,GHz. Hence, we infer that the positional accuracy of the radio-jets are only limited by their respective positional uncertainties, which in the case of R1 for example is $\sim 0.1^{\prime\prime}$.

\subsection{Tracing the sources of the jets}

Besides the detection of the radio-jets, the source of the jets are also worth noting if we are to understand better the role they play in the general evolution of the cluster. We have listed the probable location of the source galaxy for each of the jets in the final column of Table~\ref{tab:jets}, simply from their respective physical proximity (Fig.~\ref{fig:clj_core}). Starting with the assembling BCG described in detail in Sec.~\ref{sec:clj1449}, it is likely that R1 is originating from one of its members. This could in fact be the progenitor of a possible central radio AGN that is observed in many z=0 clusters. Moreover, there is another faint detection below R1 which could simply be the remnant of a previous phase of jet emission. However, due to the very low significance ($<2.5\sigma$) and large distance from any of the known cluster galaxies, we have not included it in our analysis. 

Moving slightly to the south-east, we have the R7 group surrounding N7 and S7, which is a merging pair with the latter being an X-ray detected AGN \citep{campisi09}. Hence it is highly likely that they are hotspots from the jets emerging from S7 with a level of bending that is common for AGNs interacting with the dense medium of cluster cores. Similarly, R2 could be expected to be stemming from A2. On the other hand, for the source R3, although lying within the cluster core, we can only speculate about a yet undetected highly obscured or a `switched-off' AGN in a very faint galaxy producing it.

\subsection{Radio emission: How much do we really have?} \label{sec:rad_emission_quant}

To corroborate the radio-jet conclusion, it is important to understand if there is an overall radio excess and therefore an abundance of general AGN-jet activity within \Cl. Since we are also aiming to investigate the differences in the mode of operation of radio-jets at higher redshifts compared to their low redshift counterparts, we make a comparison to the results presented in \citet{mittal09}. Here they use the X-ray flux-limited HIFLUGCS sample \citep{reiprich02} with a mean $\rm \left<z \right> \sim 0.05$ in an attempt to understand the coupling between radio-jet based AGN activity and the surrounding hot ICM. For this they measure the total radio luminosity (10\,MHz -- 10\,GHz) from the radio-jets of the centrally placed AGN of each cluster. However in \Cl, there seems to be multiple sites of AGN jet emission, the implications of which are discussed in the following section. To make a comparison, we add up the contributions of all the detected jets in place of the jets from a central AGN, which gives us a flux of $30.6\pm3.3 \rm\, \mu Jy$ and a corresponding integrated radio luminosity (10 MHz--10 GHz rest-frame) of $2.1 ~(\pm 0.2)~ \times 10^{41} ~\rm ergs~ s^{-1}$. 

It is worth noting however that this luminosity calculation is strongly dependent on the radio spectral index which can be anywhere between --0.7 \citep[as expected in case of star formation;][]{delhaize17} to --1.5 (the upper limit we calculate based on our GMRT 325\,MHz measurements). However, to use a spectral index within this range, we decide to borrow the results of \citet{hovatta14}. With a robust sample of 190 extragalactic radio-jets, they find a mean spectral index of $-1.04 \pm 0.03$ that we use in our calculation. This is also extremely close to the average spectral index of $-1$ measured by \citet{mittal09} for their sample. Although, it is worth noting that the limits of --0.7 and --1.5 result in integrated luminosities of $0.8~\times 10^{41} ~\rm ergs~ s^{-1}$ and $15.3~\times 10^{41} ~\rm ergs~ s^{-1}$ respectively. We can also have flux boosting contributions and we estimate a generous upper limit of the resulting overestimation of the jet flux to be $\sim40\%$, by assuming an additional noise contribution of $1.5\sigma$ for each source. However, accounting for this possible contribution does not change the conclusions of this paper.

\begin{figure}
    \centering
    \hspace*{-1mm}\includegraphics[width=0.48\textwidth]{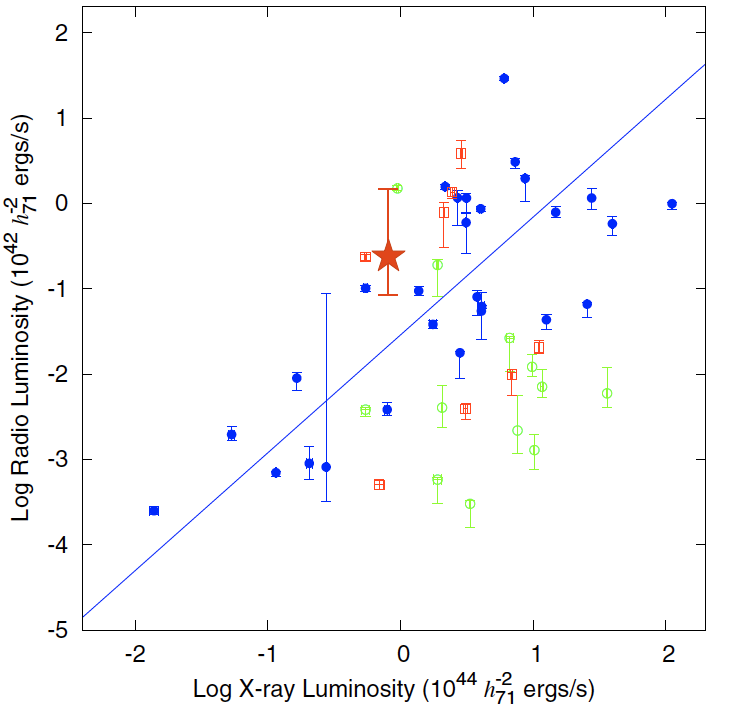}
    \caption{The total radio luminosity from radio-jets vs X-ray luminosity of the ICM (without any AGN X-ray contribution) plot from \citet{mittal09} for strong cool-core ($\rm t_{cool} < 1$ Gyr; filled blue circles), weak cool-core (1 Gyr $<\rm t_{cool} < 7.7$ Gyr; open green circles) and non cool-core ($\rm t_{cool} > 7$ Gyr; open red squares) clusters, with $\rm t_{cool}$ representing their central cooling times. \Cl\ has been represented here with the red `$\star$'. The observational error bars denote the systematic variation introduced with the change of the $\alpha$ used in the calculation of the radio luminosity. We use 0.7 (the average for star formation) as the lower limit, while for the upper limit we have 1.5 (limit estimated using the 325 MHz GMRT observation).}
    \label{fig:radio_xray}
\end{figure}

Finally, with the additional information of the X-ray luminosity ($7.2 ~ \times 10^{43} ~\rm ergs~ s^{-1}$) at 0.1--2.4 keV from the ICM of \Cl\ \citep{gobat11, valentino16}, we place this cluster among the other galaxy clusters from the \citet{mittal09} HIFLUGCS sample in Fig.~\ref{fig:radio_xray}. However the caveats of this comparison should be borne in mind. The HIFLUGCS $\rm \left<z \right> \sim 0.05$ sample occupies an entirely different epoch and evolutionary status relative to \Cl\ at z = 1.99. Differing from the low redshift clusters dominated by hot ICM plasma, clusters at $\rm z\sim2$ are rather just transitioning into this phase from an epoch featuring inflow of filamentary cold gas penetrating deep into the hot ICM \citep{valentino15,overzier16}. This difference is crucial since AGN-jet activity is known to be coupled with the level of accretion the central BH experiences. Moreover, \citet{mittal09} employed a cut-off distance between the ICM X-ray peak and the radio BCG based on a study by \citet{edwards07} which suggested that only those BCGs that lie within $70 \rm h^{-1}_{71}$ kpc of the X-ray peak of a cooling flow cluster have significant line emission and therefore enough activity to counteract the cooling. Although it is uncertain whether this conclusion is still valid at $\rm z\sim2$, we do point out that the radio-jets detected in \Cl\ are concentrated within a region of $70 \rm h^{-1}_{71}$ which includes the assembling BCG. However, we can only estimate a rough peak of the X-ray emission with the limited S/N, which still seems to be within this region.

These caveats also bring with them the points of interest once we make the comparison with the sample of galaxies in \citet{mittal09}. Although \Cl\ is at a very different redshift, its radio and X-ray luminosities place it at the upper end of the scatter in Fig.~\ref{fig:radio_xray}, without showing any stark disparity. One may bring up the aspect of us having integrated over the radio flux of all the observed jets. But repeating the same analysis with the central jet (R1) only brings down the luminosity by a factor of 4, keeping it well within the scatter and still slightly above the average. Moreover, the decentralised and plural nature of the jet emission in \Cl\ may be indicative of the inherent dynamics of the cluster core since these jets are probably coupled to accretion of material into the region. This would be a plausible situation for \Cl, given that it is most likely still experiencing cold gas accretion into its core \citep{valentino16}. 

However, a similar attempt at comparing \Cl\ to clusters (or protoclusters) at high redshifts is much more difficult. Firstly, emission from field radio sources as faint as those in \Cl\ ($\rm L_{1.4\,GHz} \sim 10^{24}\,W\,Hz^{-1}$) at $\rm z \sim 2$ is mainly attributed to star formation, with contributions from AGN activity being minimal and hard to disentangle \citep[for a review;][]{padovani16}. Studies of high redshift clusters with associated radio-jets have been limited to ones with high luminosity radio galaxies \citep[$\gtrsim 10^{27}\rm\,W\,Hz^{-1}$, e.g.,][]{wylezalek13,nesvadba17,noirot18,markov20}. This makes the detection of faint radio-jets within a cluster at $\rm z \sim 2$ relatively new.

\subsection{Quantifying the feedback}

The X-ray detectable `cavities' which result from AGN-jet activity usually allow us to quantify the amount of effective heating experienced by the ICM. The energy content of these cavities, $\rm E_{cav}$, is given by the sum of the product of the pressure and volume ($pV$), which is work done by the jet to create the cavity, and the internal energy of the radio lobes. This comes out to be $4pV$ under the assumption that the cavity is dominated by relativistic plasma. Taking this into consideration, dividing the $\rm E_{cav}$ by the cavity age, $\rm t_{cav}$, gives the cavity power, $\rm P_{cav}$, which can be considered as the lower limit of the total AGN jet power as it gauges only the `observable' effects of the jets.   

But such a direct calculation would not be possible in this work as no cavities were detected in the 4$\sigma$ X-ray diffuse emission \citep{valentino16}. However, an alternative method to measure the $\rm P_{cav}$ is given by its observed correlation with the radio luminosity, $\rm L_{\rm radio}$. Moreover, the use of such a relationship allows us to make measurements directly on the radio data \citep[e.g.,][]{best07}, thereby avoiding the issue of cavity detectability in shallow X-ray images (which is the case in our work). The aforementioned radio luminosity is usually measured from the jets of the central AGN of the cluster which is observed to be the source of entropy injection in nearby, evolved clusters. \Cl\ is however undergoing rapid evolution with it's central BCG yet to be assembled and lacks a central jet-hosting AGN. It rather features multiple AGN-jet sites distributed over the cluster core. We can hence use the integrated radio luminosity from all these jets ($\rm L_{\rm radio}$; 10 MHz--10 GHz rest-frame) that can be related to the cavity power ($\rm P_{\rm cav}$) through the following relation \citep{osullivan11}:

\begin{center}
\begin{equation}
\rm log\ P_{\rm cav} =0.71 \left (\pm 0.11 \right)\rm log\ L_{\rm radio} + 2.54 \left (\pm 0.21 \right ),\label{eq:pcav}
\end{equation}
\par\end{center}
where $\rm L_{\rm radio}$ and $\rm P_{\rm cav}$ are in units of $10^{42}\ \rm ergs \ s^{-1}$. However, one should bear in mind that this relation is still affected by uncertainties due to the assumption of $\rm E_{cav}=$ $4pV$ and the detectability of cavities within the sample used in \citet{osullivan11} amongst others, as discussed in their work. 

The $\rm P_{\rm cav}$ we estimate for \Cl\ is also dependant on the radio spectral index, $\alpha$, of the sources. Keeping in mind the caveats discussed in the previous section, we use $\alpha=-1.04$ \citep[from ][]{hovatta14} to get $\rm L_{\rm radio}$ = $2.1 ~(\pm 0.2)~ \times 10^{41} ~\rm ergs~ s^{-1}$. We put this in Eq.~\ref{eq:pcav} to get a $\rm P_{\rm cav}=1.14 ~(\pm 0.55)~ \times 10^{44} ~\rm ergs~ s^{-1}$. This power is $\sim 0.25$ times the already estimated energy injection rate calculated by \citet{valentino16} from SFR and AGN outflows ($\sim 5~\times 10^{44} ~\rm ergs~ s^{-1}$). This is a significant addition to the already abundant energy injection being experienced by the core of \Cl\ and would likely also contribute to the $\rm Ly\alpha$ halo through shocks and instabilities \citep{valentino16,daddi20} besides contributing to the overall entropy of the cluster ICM. 

Furthermore, one should also consider the age of the electron population within the jets, which determine the steepness of the spectrum. Although a determination of the age is not possible without a better sampling of the radio spectrum, the fact that we have multiple emission makes the likelihood of the jet activity to have just been `switched-on' negligible. This could in fact result in a range of spectral indices up to the --1.5 limit we have determined from the 325\,MHz data, for the different sources. The details of this distribution, although indeterminable with the current data, have large implications especially since the --1.5 limit results in a $\rm P_{\rm cav} \sim 4.7 ~ \times 10^{44} ~\rm ergs~ s^{-1}$. This is an increase of more than a factor of 4 and would make the AGN-jet feedback almost equal to the rest of the instantaneous energy injection within \Cl. This is especially important since steep spectra ($\alpha > 1$) are more likely in dense environments due to confinement of the relativistic plasma by the inter-galactic environment, which also prevents fading radio-jets from dissipating quickly \citep[e.g.,][]{giacintucci11}. This makes the aforementioned possibility of multiple jets with a range of (high) spectral indices co-existing even more likely. However, the upper limit of $-1.5$ may be too high a spectral index based on the compactness of the radio sources ($\lesssim 10$\,kpc, Fig.~\ref{fig:clj_core}). A more reasonable upper limit is possibly closer to --$1.2$ \citep[using the age-size relation presented in][]{carilli91}. This gives a $\rm P_{\rm cav} \sim 1.8 ~ \times 10^{44} ~\rm ergs~ s^{-1}$.

\subsection{Effect of low cluster mass} An additional caveat to the comparison of \Cl\ to other mature clusters with AGN-jet based energy injection (Sec.~\ref{sec:rad_emission_quant}) is its mass of $5$--$7\times 10^{13}~\rm M_{\odot}$ \citep{valentino16}, which places it rather in the galaxy group regime. \citet{giodini10} demonstrated that the mechanical energy from jets is comparable to the binding energy in galaxy groups, while it is lower by a factor of $\sim 10^2$--$10^{3}$ in clusters. This makes jets sufficient in unbinding significant fractions of the intragoup medium. Although the measurement of the binding energy is not possible in \Cl\ due to a low significance detection of the ICM X-ray emission, its galaxy group-like mass could suggest that we are underestimating the consequences of the detected radio-jets on the overall evolution of \Cl. Hence, deeper X-ray future observations of the ICM will prove crucial.   

\subsection{Galaxy 13: a possible contributor}

During the analysis of the jet driven feedback in \Cl, we have ignored galaxy 13 due to a lack of evident jet-like emission. However, it does showcase a high 3\,GHz radio flux ($6.6\pm1.3\,\mu$Jy) only two-thirds of which can be accounted for by the intrinsic SFR. The rest may still suggest a level of kinetic feedback present in this galaxy with the jets having not been detected due to projection effects. Assuming this to be true, this would only lead to an increment of $\sim 6\%$ in the total jet power. Although, this is a small contribution, it is very important to note that this galaxy represents an additional site of AGN activity (with or without jets) which has crucial implications on our understanding of \Cl, as described in the conclusion of this work.

\subsection{The SZ residual}

\begin{figure}
    \centering
    \hspace*{-1mm}\includegraphics[width=0.48\textwidth]{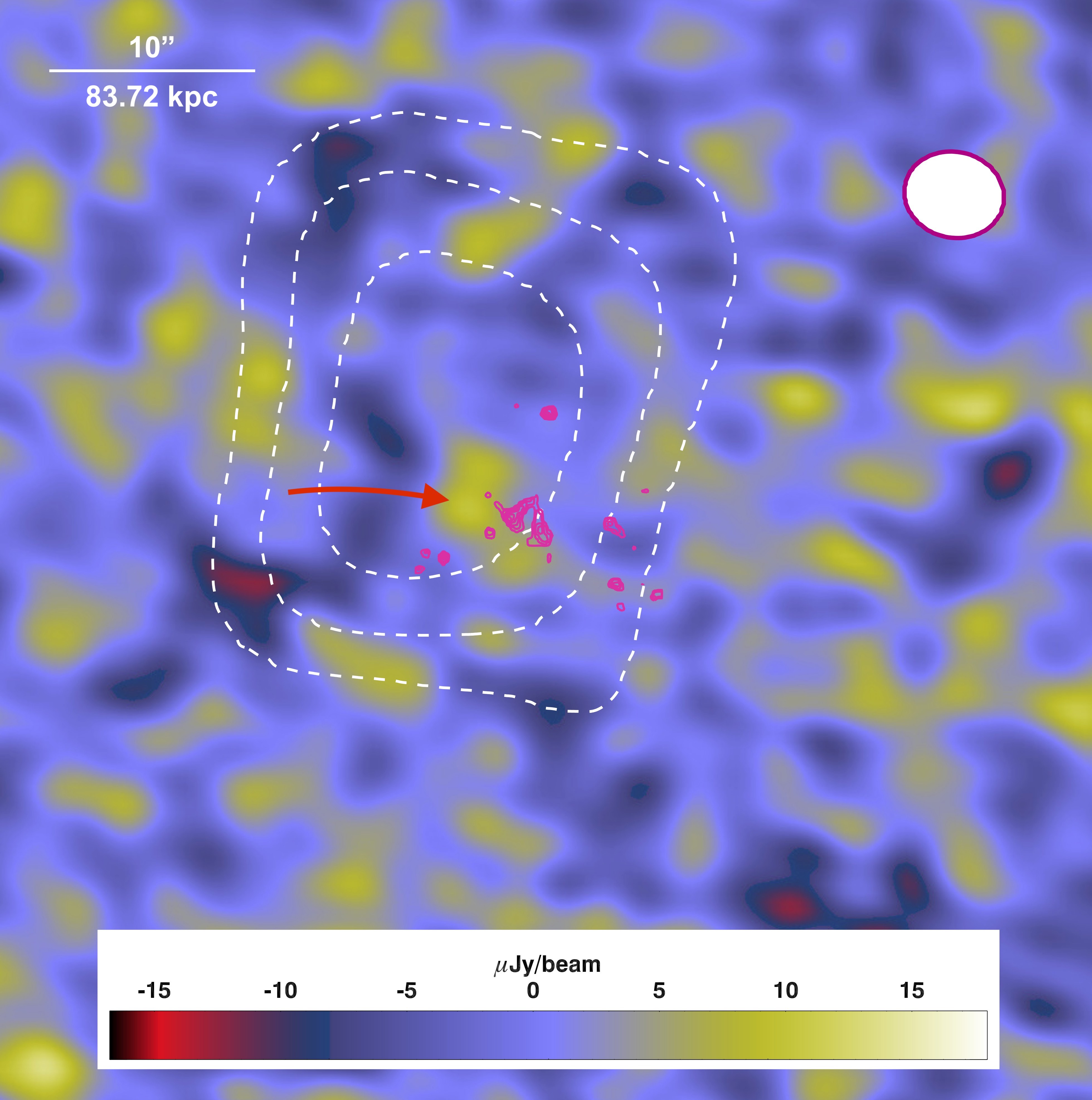}
    \caption{ALMA 92 GHz residual image of \Cl\ borrowed from \citet{gobat19}, after subtracting both the point sources and the SZ signal from the cluster (denoted by the white dashed contours). The colour bar at the bottom denotes the residual flux per beam with a positive signal indicating a lack of gas density and vice-versa. This has been overlaid with the radio contours in magenta while the white-filled red ellipse shows the size of the synthesised ALMA beam. The red arrow points to the possible under-density due to the jets.}
    \label{fig:sz_res}
\end{figure}

Whenever strong radio-jets occur within dense cluster cores, resulting under-densities are observed due to the gas in the ICM being blown out by the kinetic feedback to form cavities. These cavities are usually spatially coincident with the lobes of the radio jet and can be observed in X-ray \citep[for a review, ][]{fabian12}. But with the diffuse X-ray ICM emission having been observed at only 4$\sigma$ \citep{valentino16}, it is not possible to detect variations indicative of cavities. However, in case there are under-densities, they should also manifest in the thermal-SZ map of the cluster as positive signals due to a lack of the SZ decrement \citep[that is otherwise expected from the dense ICM gas; ][]{komatsu01} in the 92\,GHz flux at z=2. \citet{gobat19} presented a detailed study of this for \Cl\ and also generated a residual map after the subtraction of the 5$\sigma$ SZ signal of the ICM. In Fig.~\ref{fig:sz_res} we superimpose this map with the location of the radio-jets. Interestingly, there seems to be a hint of an under-density (in the form of a positive signal), albeit of a 2$\sigma$ significance, between sources R1 and R7\_W which are two of the brightest sources and therefore strongest jets in our list. This is suggestive of a marginal decrease in the electron density due to possible presence of jet driven feedback, in line with our conclusions.  

\subsection{Inverse-Compton contribution}

Following the detection of radio-jets within \Cl, we revisit the detection of diffuse X-ray emission in the cluster \citep[luminosity at 0.1--2.4 keV $= 7.2 ~ \times 10^{43} ~\rm ergs~ s^{-1}$; ][]{gobat11, valentino16}. This has been attributed to the hot ICM of the cluster. However, it is now imperative to check any possible contributions from the newly detected jets to the cluster X-ray flux through inverse-Compton (IC) scattering \citep[][ for a review]{worrall09}. The cause of this is the relativistic electron population (already emitting in radio from the jet sites) upscattering the photon-field primarily from the Cosmic Microwave Background. \citet{fabian09} demonstrated that radio emission corresponding to IC is best detected at lower frequencies, as higher frequencies like 3\,GHz may not be sensitive enough to aging electron populations with steep spectra. Hence we decide to only use the GMRT 325\,MHz upper limit for this estimation. Using the equations (4.53) and (4.54) in \citet{tucker75}, we get an upper limit IC contribution of $6.5~ \times 10^{42}~ \rm ergs~s^{-1}$ ($\sim 10\%$ of the X-ray luminosity). This is measured using a Lorentz factor of $10^{3}$--$10^{5}$ and a photon index of 2, the reasons for which has been provided in \citet{fabian09} who estimate the IC contributions for HDF130 at z=1.99: a case very similar to ours. 

Furthermore, we stack the \textit{Chandra} X-ray data at the locations of the jets to observationally confirm a lack of IC X-ray emission. The generous upper limit hence derived from the total counts detected is $2.2~ \times 10^{42}~ \rm ergs~s^{-1}$ or $3\%$ of the ICM X-ray luminosity. We do not include the R7 group in this analysis though, in order to prevent contamination from the X-ray flux of the core of the AGN in the adjacent S7. However, we are certain that we do not miss any contribution to the ICM X-ray luminosity since its value of $7.2 ~ \times 10^{43} ~\rm ergs~ s^{-1}$ was calculated after subtracting the flux from S7 \citep{gobat13, valentino16}. Hence, even if there is minimal contribution from the R7 jets, it has already been excluded. Therefore, we conclude that there is no major IC contribution to the cluster X-ray luminosity, which has already been found to trace the cluster mass due to an agreement with the SZ measurements presented in \citet{gobat19}.

\section{Summary and Conclusion} \label{sec:conclusion}

We have presented a detailed analysis of the radio behaviour of \Cl, motivated by a revised reduction of 3\,GHz VLA observations as well as new 325\,MHz GMRT data. The main results of this study are as follows:

\begin{enumerate}
    \item We reach a sensitivity of $1.34\, \mu$Jy/beam at 3\,GHz, which is a $\sim$34$\%$ improvement compared to the previous results presented in \citet{coogan18}. This improved sensitivity allowed us to detect multiple radio emission regions without any counterparts in our \textit{HST/WFC3} NIR as well as $870\,\mu$m continuum data. We also use median stacking to check for low levels of star formation or stellar populations at these sites that would result in faint emissions in sub-mm or NIR respectively, and find none within the detection limits.
    Finally, the possible association with known galaxies due to physical proximity for 4/6 of these objects without any overlap led us to conclude them to be AGN radio-jets.  
    
    \item Detection of multiple radio-jet sites in a cluster core is in stark contrast to low redshift counterparts, which predominantly feature centrally placed radio AGN-jets. We furthermore measure a total flux of $30.6 \pm 3.3\,\mu$Jy from all the detected jets. Using a spectral index of --1.04 based on the study by \citet{hovatta14} for radio-jets, we determine an integrated radio luminosity (10\,MHz -- 10\,GHz) of $2.1 ~(\pm 0.2)~ \times 10^{41} ~\rm ergs~ s^{-1}$ which places \Cl\ at the radio-bright end of the radio-jets to X-ray luminosity relation for low redshift clusters (Fig.~\ref{fig:radio_xray}). Additionally, given that the radio luminosity is measured only from the central dominant AGN-jets in \citet{hovatta14}, we point out that using only the jet detected at the site of the BCG assembly (R1) for this analysis does not change this conclusion.   
    
    \item With the spectral index of --1.04, we further estimate that $\sim 20\%$ of the total instantaneous energy injection into the ICM of \Cl\ can be attributed to AGN jets, with a $\rm P_{\rm cav}=1.14 ~(\pm 0.55)~ \times 10^{44} ~\rm ergs~ s^{-1}$. Although, this could be much higher (up to $4.7~ \times 10^{44} ~\rm ergs~ s^{-1}$) given that the multiple sites suggest an extended period of AGN-jet activity which could lead to a spectral index even up to the limit of --1.5 measured from our 325\,MHz GMRT data. However, a more realistic upper limit is found to be closer to --1.2, which gives $\rm P_{\rm cav} \sim 1.8 ~\times 10^{44} ~\rm ergs~ s^{-1}$. Only deeper low-frequency observations can settle this uncertainty. 

    \item We make 3\,GHz flux measurements to estimate the TIR-radio luminosity ratio ($\rm q_{TIR}$) of the galaxies within the cluster core. We find agreement within 0.2 dex with the results of \citet{delvecchio20} for star-forming main sequence galaxies.

\end{enumerate}

Based on this work, we hence expand our current understanding of \Cl. We begin with the revised 3\,GHz flux measurements and the subsequent FIR-radio correlation analysis. We find the cluster galaxies to be in agreement with the relation reported in \citet{delvecchio20}, hence suggesting that galaxies in clusters follow the same FIR-radio correlation. This conclusion also reinforces the use of radio luminosity as an SFR tracer even in dense environments. However, as it is beyond the scope of this work, studies of a larger sample of clusters would be highly encouraged to understand more subtle effects of density on this relation.

Moving on to the detection of multiple sites of AGN jet-like emission which has been the primary focus of this work, we revise our narrative of the cluster galaxies. It is worth noting that \Cl\ showcases an SFR $\rm \sim 1000~ M_{\odot}yr^{-1}$ at its core, which is most likely being driven by residual cold-gas accretion into the cluster core \citep{valentino16}. The same accretion could very well be resulting in the presence of the two radiative-mode radio-detected AGNs in the cluster core, at least one of which is featuring clear jets, since they are known to be present in environments with high levels of accretion \citep{lin10} and star formation \citep{kauffmann03,kauffmann09}. The situation at the site of the BCG assembly is a little more obscure however. The jet R1 may be stemming from the two massive quiescent bulge-dominated galaxies (B1 and H5) or the highly star-forming A1 which could be hosting a highly obscured AGN core. However in this case as well, the cause of AGN activity can be linked to accretion, especially since the same accretion is most likely driving the star formation in A1 \citep{valentino16}. Although, one cannot dismiss an additional contribution from merger activity in \Cl\ \citep{coogan18} towards the triggering of AGN-jet activity. This is especially important since both the assembling BCG and the N7-S7 pair are known to be undergoing mergers and are the sites of R1 and the R7 group respectively. 

Regarding the effect of the jets on the hot ICM of \Cl, we revisit the results of \citet{valentino16} that were motivated by the discovery of a giant 100\,kpc extended $\rm Ly\alpha$ nebula in the cluster core. They found the AGN outflows from the two radiative mode AGNs (13 and S7) to be the most likely candidates powering the nebula, thereby disfavouring an established classical cooling flow in the cluster core. Adding to this is the presence of multiple sites of AGN-jet driven energy injection, which has led us to conclude the presence of a dispersed mode of AGN kinetic feedback in \Cl. Since the centralised AGN driven feedback at the cores of low redshift CC clusters is crucial for balancing the global cooling \citep[][for a review]{mcnamara07}, the peak of which is usually coincident with the AGN itself \citep{edwards07}, the lack of such an arrangement would suggest an absence of a classical cooling flow. This is hence in line with the conclusion of \citet{valentino16}. Moreover, it is highly likely that the jets are also contributing to the powering of the $\rm Ly\alpha$ nebula which is already too high in luminosity (by a factor of 10--1000) for it to be present in a CC cluster. Another conclusion that can be drawn is the improbability of a change in the status-quo with any phases of cooling flow domination in the immediate future of \Cl\ during an absence of AGN activity, since this is unlikely to be allowed given the multiple AGN sites. Hence, we are possibly witnessing a `steady state' of the cluster ICM due to a rather constant version of the AGN feedback driven loop \citep{mcnamara05, rafferty06, gaspari17}. 

The lack of a classical cooling flow however should not be interpreted as a sign of low overall AGN-jet driven feedback in \Cl. We demonstrate that the ratio of the radio luminosity from jets and the ICM X-ray luminosity is higher than what is usually expected even in CC clusters (Fig.~\ref{fig:radio_xray}). This is especially crucial considering the sizable contribution of the jets to the total energy injection rate in this cluster which we estimate to be $\sim 20\%$, but could in fact be even higher if one considers possible effects of an older electron population. It is unclear although if this puts \Cl\ on a fast track to becoming a non cool-core cluster \citep[with $\rm t_{cool}>7Gyr$ at the cluster core, Fig.~\ref{fig:radio_xray}; ][]{mittal09} or could it still end up turning into a CC cluster with a central dominant AGN spewing out relativistic jets and harboring a classical cooling flow at later stages. Uncovering evolutionary pathways, if they exist, would require a larger sample of galaxy clusters at high redshifts with detailed study of the interaction between jets and their ICM. Deeper ALMA SZ observations tracing the morphology of the ICM of clusters could facilitate such a study in the future. However, a clear picture would most likely only be possible in the 2030's with the coming of ATHENA and with it an era of high sensitivity X-ray observations. Until then, it is imperative that we keep attempting similar studies as that done for \Cl\ to pave the way for future efforts.

\section*{Acknowledgements}
First and foremost, we would like to thank the anonymous referee for the valuable suggestions. This paper makes use of JVLA program 12A-188. The National Radio Astronomy Observatory is a facility of the National Science Foundation operated under cooperative agreement by Associated Universities, Inc. Also used are data from the GMRT project 23\_068, for which we thank the staff of the GMRT who made the observations possible. GMRT is run by the National Centre for
Radio Astrophysics of the Tata Institute of Fundamental
Research. 

Moreover, this paper makes use of the following ALMA data: 2012.1.00885.S and 2015.1.01355.S. ALMA is a partnership of ESO (representing its member states), NSF (USA) and NINS (Japan), together with NRC (Canada), MOST and ASIAA (Taiwan), and KASI (Republic of Korea), in cooperation with the Republic of Chile. The Joint ALMA Observatory is operated by ESO, AUI/NRAO and NAOJ. 

V.S. acknowledges the support from the ERC-StG ClustersXCosmo grant agreement 716762. B.S.K. would like to thank Chiara D'Eugenio (CEA-Saclay) and Mengyuan Xiao (CEA-Saclay) for the endless discussions that fuelled the completion of this work. Finally, B.S.K. would also like to express his gratitude to Francesco Carotenuto (CEA-Saclay) for his valuable suggestions regarding the radio interferometry data reduction involved in this work. 

\section*{Data Availability}

The data underlying this article are publicly available in the online archives of VLA, GMRT and ALMA. They can be accessed under their respective project codes: 12A-188 (VLA), 23\_068 (GMRT), 2015.1.01355.S (band 4 ALMA) and 2012.1.00885 (band 3 and 7 ALMA).  



\bibliographystyle{mnras}

%\begin{thebibliography}{99}
%\end{thebibliography}

%%%%%%%%%%%%%%%%%%%%%%%%%%%%%%%%%%%%%%%%%%%%%%%%%%

%%%%%%%%%%%%%%%%% APPENDICES %%%%%%%%%%%%%%%%%%%%%

%\appendix



%%%%%%%%%%%%%%%%%%%%%%%%%%%%%%%%%%%%%%%%%%%%%%%%%%

% Don't change these lines
\bsp	% typesetting comment
\label{lastpage}
\end{document}